\documentclass[11pt,a4paper]{article} 
 \usepackage[numbers,sort&compress]{natbib}
 \usepackage{amsmath}
\usepackage{amssymb}
\usepackage{authblk}
\usepackage{graphicx}
\usepackage{comment}
\usepackage{braket}
\usepackage{caption}
\usepackage{multirow}
\usepackage[toc,page]{appendix}
\usepackage{amsfonts}
\usepackage{rotating}

\title{Experimental demonstration of nonbilocality with truly independent sources and strict locality constraints}

\author[1,2]{Qi-Chao Sun}
\author[1,2]{Yang-Fan Jiang}
\author[1,2]{Bing Bai}
\author[3]{Weijun Zhang}
\author[3]{Hao Li}
\author[1,2]{Xiao Jiang}
\author[1,2]{Jun Zhang}
\author[3]{Lixing You}
\author[4]{Xianfeng Chen}
\author[3]{Zhen Wang}
\author[1,2]{Qiang Zhang}
\author[1,2]{Jingyun Fan}
\author[1,2]{Jian-Wei Pan}

\affil[1]{  National Laboratory for Physical Sciences at Microscale and Department of Modern Physics, Shanghai Branch, University of Science and Technology of China, Hefei, Anhui 230026, China}
\affil[2]{CAS Center for Excellence and Synergetic Innovation Center in Quantum Information and Quantum Physics, Shanghai Branch, University of Science and Technology of China, Hefei, Anhui 230026, China}
\affil[3]{ State Key Laboratory of Functional Materials for Informatics, Shanghai Institute of Microsystem and Information Technology, Chinese Academy of Sciences, Shanghai 200050, China}
\affil[4]{ Department of Physics and Astronomy, Shanghai Jiao Tong University, Shanghai, 200240, China}

\date{\today}

\begin{document}

\maketitle

\textbf{Entanglement swapping entangles two particles that have never interacted~\cite{ZukowskiEntanglementSwapping}, which implicitly assumes that each particle carries an independent local hidden variable, i.e., the presence of bilocality~\cite{Branciard2010Bilocal}. Previous experimental studies of bilocal hidden variable models did not fulfil the central requirement that the assumed two local hidden variable models must be mutually independent~\cite{Carvacho2016BellNetwork,Saunders2017Bilocal,Andreoli2017Fram}. By harnessing the laser phase randomization~\cite{Yuan2014PhaRand} rising from the spontaneous emission to the stimulated emission to ensure the independence between entangled photon-pairs created at separate sources and separating relevant events spacelike to satisfy the no-signaling condition, for the first time, we simultaneously close the loopholes of independent source, locality and measurement independence in an entanglement swapping experiment in a network. We measure a bilocal parameter of $1.181\pm0.004$ and the CHSH game value of $2.652\pm0.059$, indicating the rejection of bilocal hidden variable models by $45$ standard deviations and local hidden variable models by $11$ standard deviations. We hence rule out local realism and justify the presence of quantum nonlocality in our network experiment. Our experimental realization constitutes a fundamental block for a large quantum network. Furthermore, we anticipate that it may stimulate novel information processing applications~\cite{Chaves2016BellNetwork,Lee2018DIInfProc}.}

Quantum nonlocality is incompatible with local realism~\cite{EPR,Bell1964}. Several recent Bell test experiments have made remarkable progress to disprove (single) local hidden variable models by closing the detection and locality loopholes simultaneously~\cite{Hensen2015LoopholeFreeBT,Giustina2015LoopholeFreeBT,Shalm2015LoopholeFreeBT,Rosenfeld2016Bell}. Quantum nonlocality is a resource for many information processing tasks, particularly for device-independent quantum information processing applications~\cite{Colbeck2007Thesis,Acin2006BellQKD,2014VaziraniDI}. Disproving multi-local hidden variable models in quantum networks is part of the campaign against local realism\cite{Branciard2010Bilocal,Tavakoli2014StarNet,Chaves2016BellNetwork,Rosset2016Nlocal,Tavakoli2017StarNet,Carvacho2016BellNetwork,Saunders2017Bilocal}, which not only justifies the nonlocality as the baisc property of the quantum network, but also holds potential for novel quantum information processing applications~\cite{Chaves2016BellNetwork,Lee2018DIInfProc}.

\begin{figure}
\centering
\includegraphics[width=0.6\textwidth]{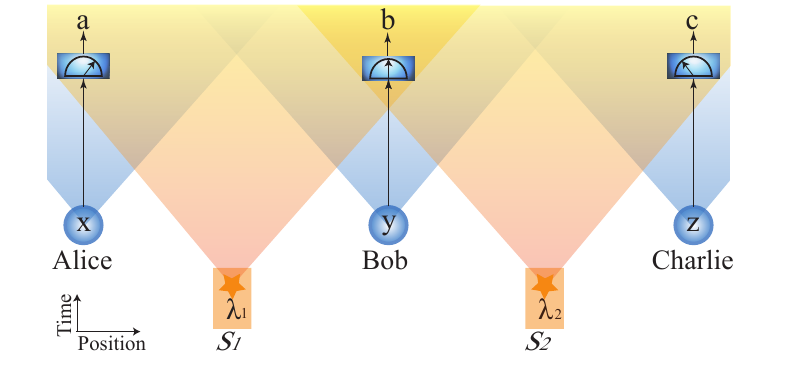}
\caption{\textbf{The space-time diagram of the simplest quantum network with two sources distributing entanglement to three nodes with shaded area indicating light cones. }}
\label{fig:scheme}
\end{figure}

Considering the simplest quantum network shown in Fig.~\ref{fig:scheme}, source $S_1$ distributes the Bell state $\ket{\Phi^+}_{AB}=\frac{1}{\sqrt{2}}(\ket{H}_A\ket{H}_B+\ket{V}_A\ket{V}_B)$ between Alice and Bob and source $S_2$ distributes the Bell state $\ket{\Phi^+}_{B^\prime C}=\frac{1}{\sqrt{2}}(\ket{H}_{B^\prime}\ket{H}_C+\ket{V}_{B^\prime}\ket{V}_C)$ between Bob and Charlie, where $\ket{H}$ and $\ket{V}$ denote the horizontal and vertical polarization quantum state, respectively. Entanglement swapping is realized conditioning on the Bell state measurement (BSM) by Bob. The particles held by Alice and Charlie which have \emph{never interacted before} become entangled~\cite{ZukowskiEntanglementSwapping}. Noting that the recent loophole free experimental realizations of violating the Bell inequality rely on an assumption (or a similar one) that a local hidden variable is created along with the birth of a state in the source~\cite{Giustina2015LoopholeFreeBT,Shalm2015LoopholeFreeBT}. Entanglement swapping relies on an assumption that the two state creation events at the two sources are mutually independent; each event is independently assigned a local hidden variable to carry the exact state information, i.e., $\lambda_1$ is created in source $S_1$ and passed to Alice and Bob, and $\lambda_2$ is created in source $S_2$ and passed to Bob and Charlie.
According to local hidden variable theories, the measurement outcomes $a$, $b$, and $c$ of Alice, Bob, and Charlie at the three nodes are completely predetermined for measurement setting choices $x$, $y$, $z$ and local hidden variables $\lambda_1$ and $\lambda_2$, respectively, such as $a=a(x,\lambda_1)$, $b=b(y,\lambda_1,\lambda_2)$, and $c=c(z,\lambda_2)$. The tripartite probability distribution under the bilocal hidden variable assumption may be given by
\begin{equation}
P(a,b,c|x,y,z)=\int\mathrm{d}\lambda_1\mathrm{d}\lambda_2\rho(\lambda_1,\lambda_2)P(a|x,\lambda_1)P(b|y,\lambda_1,\lambda_2)P(c|z,\lambda_2),
\label{eq:bilocal}
\end{equation}
where $P(a|x,\lambda_1)$, $P(b|y,\lambda_1,\lambda_2)$, and $P(c|z,\lambda_2)$ are probabilities of local measurements at the three nodes, respectively. The independent sources and locality condition require the probability distribution of local hidden variables to be factorable, $\rho(\lambda_1,\lambda_2)=\rho(\lambda_1)\rho(\lambda_2)$ with $\int\mathrm{d}\lambda_1\rho(\lambda_1)=1$ and $\int\mathrm{d}\lambda_2\rho(\lambda_2)=1$. Branciard et al. proved that some correlation functions can be used to test the bilocal hidden variable models in the similar spirit of the Bell inequality~\cite{Bell1964,Branciard2010Bilocal,Branciard2012Bilocal}, which appear in the form of
\begin{equation}
\mathcal{B}=\sqrt{I}+\sqrt{J},
\label{eq:B13}
\end{equation}
where $I$ and $J$ are linear combinations of tripartite probability distributions. Bilocal hidden variable models require $\mathcal{B}\le1$. $\mathcal{B}>1$ indicates the rejection of bilocal models.

The experimental tests of bilocal relations demand significantly more than the standard Bell test experiments. Particular attention has to be paid to ensure space-like separation between the relevant events of entanglement creation and detection in a quantum network which has a complicated causal structure. Yet, it still remains a challenge to create two mutually independent sources of pairs of entangled particles~\cite{Yang2006IndSource,Kaltenbaek2006IndSource,Sun2016Tel}, which may be referred to as the independent source loophole. The most recent attempt was to synchronize two spontaneous four-wave mixing (SFWM) processes to create entangled photon-pairs by modulating the pump laser pulses, in which the outputs of two continuous wave (cw) lasers were carved into pulses to drive SFWM processes~\cite{Sun2016Tel}. One might argue that a correlation was established between the two cw lasers prior to the occurrence of the two SFWM processes~\cite{Branciard2010Bilocal,Saunders2017Bilocal}. Previous experimental tests against bilocal hidden variable models in a quantum network failed both the space-like separation and independent source requirements~\cite{Carvacho2016BellNetwork,Saunders2017Bilocal,Andreoli2017Fram}.  Here we present in this Letter an experimental study to reject bilocal hidden variable models with both locality and independent source loopholes closed. Our experiment may serve as a valuable reference to the realization of the future quantum networks.

We construct a network in the Shanghai campus of University of Science and Technology of China, in which sources $S_1$ and $S_2$ distribute polarization-entangled photon-pairs to the three nodes (see Fig. 2a). Bob sandwiches polarization beam splitters (PBSs) between  50:50 beam splitters (BSs) to realize the BSM (see Fig. 2c). The setup can read out the path, polarization, and photon number information about the incoming two photons: the two photons in Bell state $\ket{\Psi^-}$ exit from different ports of the first BS, the two photons in Bell state $\ket{\Psi^+}$ exit from different ports of the PBS, the two photons in Bell state $\ket{\Phi^+}$ or $\ket{\Phi^-}$ bunch together and are resolved with $50\%$ of success by the photon-number resolving detection (implemented by the last BS and two single photon detectors). For the BSM with 1-fixed input and 3-output, we examine the bilocal relation $\mathcal{B}_{13}$ (see Supplementary Information). Alice (Charlie) randomly selects one from the two measurement settings, $A_0$ or $A_1$ ($C_0$ or $C_1$), for single photon polarization state measurement upon receiving a bit from the quantum random number generator. To maximize the value of $\mathcal{B}_{13}$, we set $\hat{A}_0=\hat{C}_0=(\sqrt{2}\sigma_z+\sigma_x)/\sqrt{3}$ and $\hat{A}_1=\hat{C}_1=(\sqrt{2}\sigma_z-\sigma_x)/\sqrt{3}$. According to quantum mechanics, we have $\mathcal{B}_{13}>1$ for quantum state with a swapped entanglement visibility $v>2/3$. As a comparison, we note that the lowest visibility required to violate the Bell inequality is $ v=1/\sqrt{2}$ . 

\begin{figure}
\centering
\includegraphics[width=\textwidth]{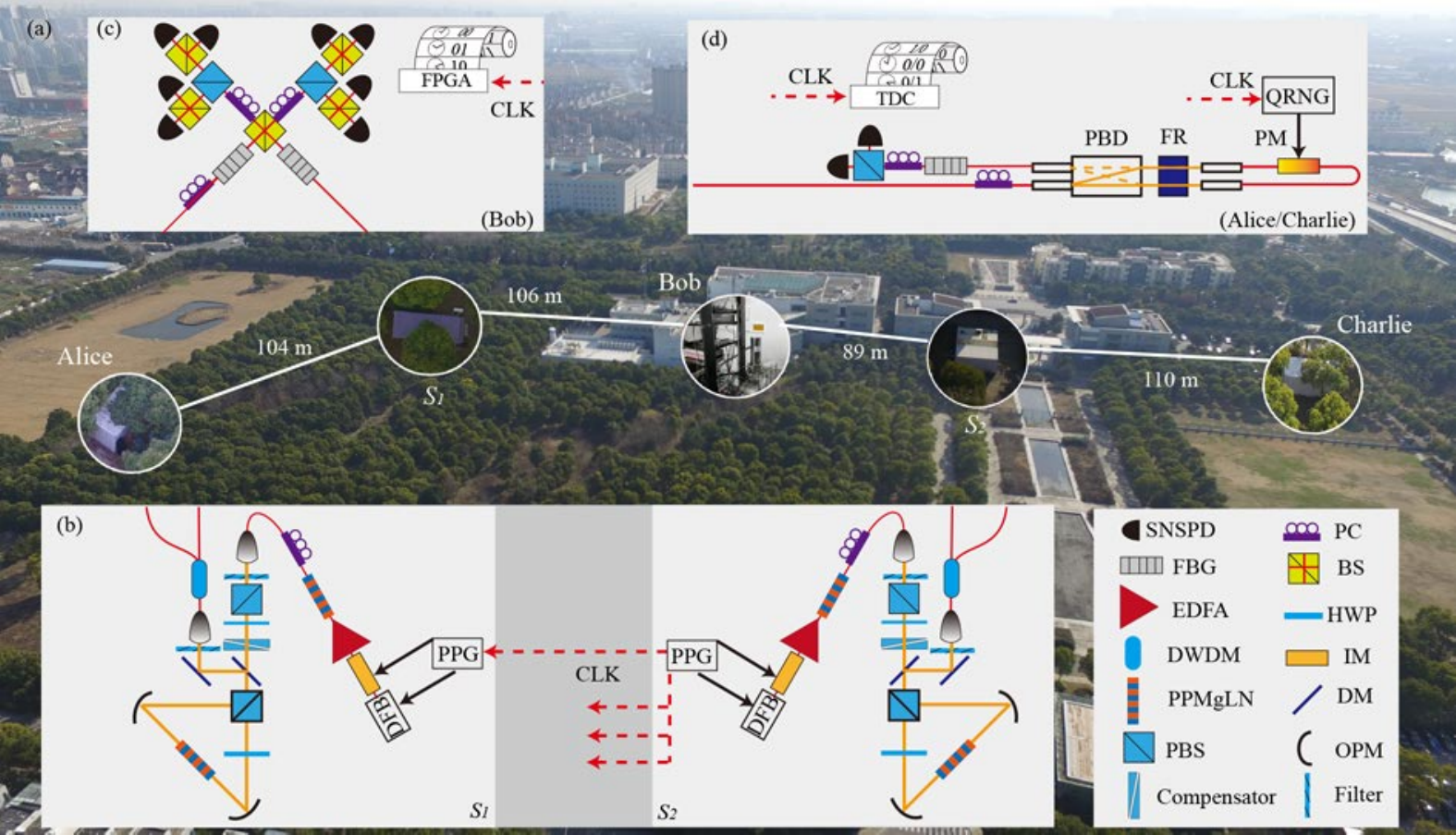}
\caption{\textbf{Schematics for testing the bilocality in a network.} \textbf{a}, The bird's-eye view of the network with nodes for Alice, Bob and Charlie and two sources $S_1$ and $S_2$. The length of the connecting fibre links (white lines) for Alice-$S_1$, $S_1$-Bob, Bob-$S_2$, and $S_2$-Charlie are 110.98 m, 124.52 m, 108.13 m, and 124.55 m, respectively. \textbf{b}, In each source, the laser pulses from an 1558-nm DFB laser is frequency-doubled in a PPMgLN crystal after passing through an EDFA. The produced 779-nm laser pulses create entangled photon-pairs in the second PPMgLN crystal, which are collected into optical fibre. In $S_2$, a 12.5-GHz microwave clock (PPG) is used as the master clock which sends synchronization signals (CLK) to all nodes in the network (red dashed lines). See Supplemental Information for details about the experimental realization. \textbf{c}, Bob performs the Bell state measurement (BSM) with the combination of 50:50 beam splitters (BS),  polarizing beam splitters (PBSs), and eight superconducting nanowire single photon detectors (SNSPD). The photon-detection results are analyzed in real-time and recorded by a field programmable gate array (FPGA) which is synchronized by the CLK signals. \textbf{d}, Synchronized to the master clock, a quantum random number generator (QRNG) outputs a random bit to notify Alice (Charlie) to perform the single photon polarization modulation (measurement setting choice) with a loop interferometer, which consists of a polarization beam displacer (PBD), a Faraday rotator (FR) and a phase modulator (PM) (see text). The single photon detection results are recorded by a time-to-digital converter (TDC) which is synchronized by the CLK signals.}
\label{fig:setup}
\end{figure}

In each of the two sources $S_1$ and $S_2$, we inject 90-ps, 779-nm laser pulses at a repetition rate of 250 MHz into a Sagnac loop which contains a 2.5-cm-long, periodically poled MgO doped Lithium Niobate (PPMgLN) crystal in the middle (Fig. 2b). The loop emits polarization-entangled photon-pairs via spontaneous parametric down-conversion (SPDC) process, which are coupled into single mode optical fibre to be delivered to the 3 nodes. To suppress the distinguishability between photons from separate sources in the network, we first extract the pair of photons at the phase-matching wavelengths of 1560~nm and 1556~nm with inline dense wavelength-division multiplexing (DWDM) filters and then pass photons through inline 3.3-GHz fibre Bragg gratings (FBGs) to suppress the spectral distinguishability; the 133-ps coherence time of single photons is much longer than the pump pulse duration, which, together with the high bandwidth synchronization (with an uncertainty of 4~ps) suppress the temporal distinguishbility; the good fibre optical mode eliminates the spatial distinguishability (see Supplementary Information).

Independent sources of entanglement are central to the realization of quantum networks. Previous attempts to create independent sources are likely subject to the concern of common past. In this experiment, we switch an electrically driven laser diode from much below the threshold to well above the threshold in each duty cycle such that the phase of each generated laser pulse is randomized in each source~\cite{Yuan2014PhaRand}. The same mechanism was employed to satisfy the requirement of measurement independence in previous loophole free Bell test ~\cite{Abellan2015QRNG,Hensen2015LoopholeFreeBT,Giustina2015LoopholeFreeBT,Shalm2015LoopholeFreeBT}. The two SPDC processes in the two sources are therefore disconnected. In this way we close the independent source loophole.  Microwave clocks are used to synchronize all events in the experiment (see Supplemental Information). Receiving a signal from the microwave clock triggers the generation of a laser pulse for the creation of a pair of entangled photons via SPDC process, which also stands for the earliest time for the birth of a local hidden variable in that duty cycle.

We assign a duty cycle an experimental trial. To satisfy the requirements of measurement independence and locality constraint, we require to satisfy four space-like separation conditions in each experimental trial: (1) space-like separation between the two state emission events in sources $S_1$ and $S_2$; (2) space-like separation between the events of Alice (Charlie) completing the quantum random number generation for  measurement setting choice and the state emission events in sources $S_1$ and $S_2$; (3) space-like separation between the two events of random bit generation for measurement setting choices for Alice and Charlie; (4) space-like separation between the event of Alice (Charlie) completing the quantum random number generation for measurement setting choice and the events of  completing the single photon detections by Bob and Charlie (Alice). The space-time diagrams depicted  in Fig.~3 show that our experimental realization satisfies all of these requirements. Note that all diagrams in Fig.~3 are drawn to the scale (see Supplementary Information).

\begin{figure}
\centering
\includegraphics[width=0.8\textwidth]{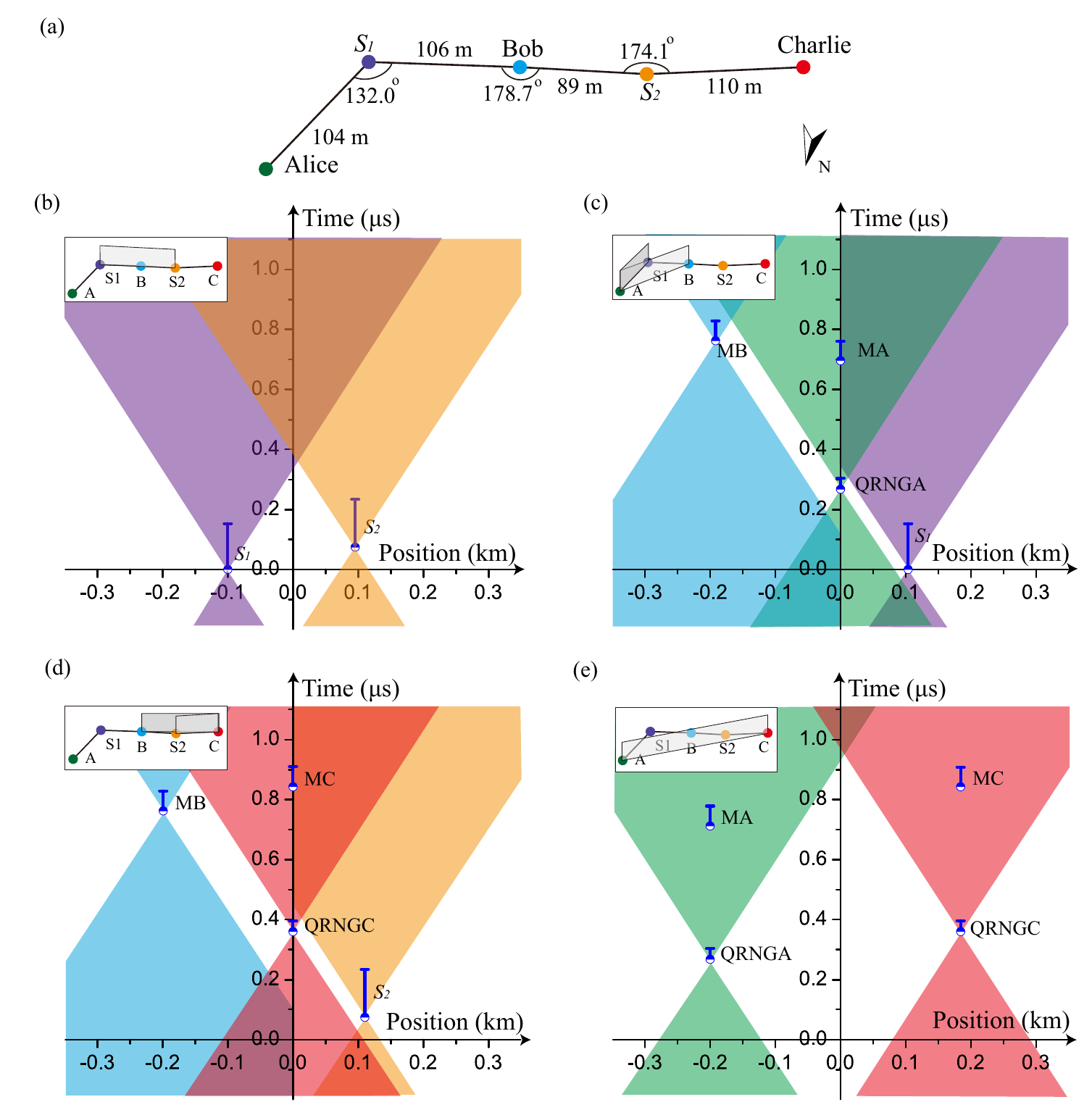}
\caption{\textbf{Space-time configuration of relevant events in each experimental trial.} \textbf{a}, Relative spatial configuration of the 2-source and 3-node network. \textbf{b}-\textbf{e}, Space-time diagrams of the relationship between important events in the nodes which is indicated in the insets. The origin of the axes are displaced to reflect the relative space and time difference between them. \textbf{b}, Space-like separation between state emission events in the two sources, $S_1$ and $S_2$.  \textbf{c}, Space-like separation between quantum random number generation event in Alice (QRNGA) and the measurement event by Bob (MB) is shown on the left hand side of the vertical axis, meanwhile,  the state emission event in the nearest source on the right hand side of the vertical axis. \text{d}, Similar with \text{c}, QRNGC denotes the quantum random number generation event of Charlie. \text{e}, Space-like separation between the two quantum random generation events and space-like separations between quantum random number generation event and measurement event between the node Alice and node Charlie.  The blue vertical bars represent the time elapses of events, with starting and ending marked by circles and horizontal dashes, respectively. (see Supplementary Information for details). }
\label{fig:spacetime}
\end{figure}

A high speed high fidelity single photon polarization modulation device is a critical element in the realization of the quantum network. We present such an implementation based on the design of a loop interferometer. As shown in Fig.\ref{fig:setup}~(d), a single photon incident onto the loop has its two orthogonal polarization components exit at different ports of the polarizing beam dispacer (PBD). With polarization rotated by $45^o$ by the Faraday rotator (FR) and aligned with the slow axis of polarization-maintaining fibre (PMF), both polarization components are coupled into the PMF to propagate in opposite directions in the loop. A phase modulator (PM) is displaced from the middle position by 26~cm to create a relative delay of about 1.3~ns between the arrival times of the two counter-propagating components at the PM such that the PM can manipulate the phase to only one of them. The two components interfere at the PBD and exit as a single photon pulse with a modulated polarization state. We demonstrate to realize the single photon polarization state modulation at a rate of 250 MHz with random inputs with a fidelity of $(99.0\pm0.2)\%$.

As a reliability check prior to performing the experimental test of bilocal hidden variable models, we measure the two-photon interference visibility to be greater than $97\%$ for states prepared by both sources; we obtain a fitted visibility of $(96.5\pm1.6)\%$ in the Hong-Ou-Mandel measurement with photons from the two independent sources~\cite{HOM}. 
We attribute the imperfect visibility mainly to the multi-photon-pair events in the SPDC process. The four-fold coincidence count rate is about 1 per second in the experiment. 

\begin{figure}
\centering
\includegraphics[width=0.5\textwidth]{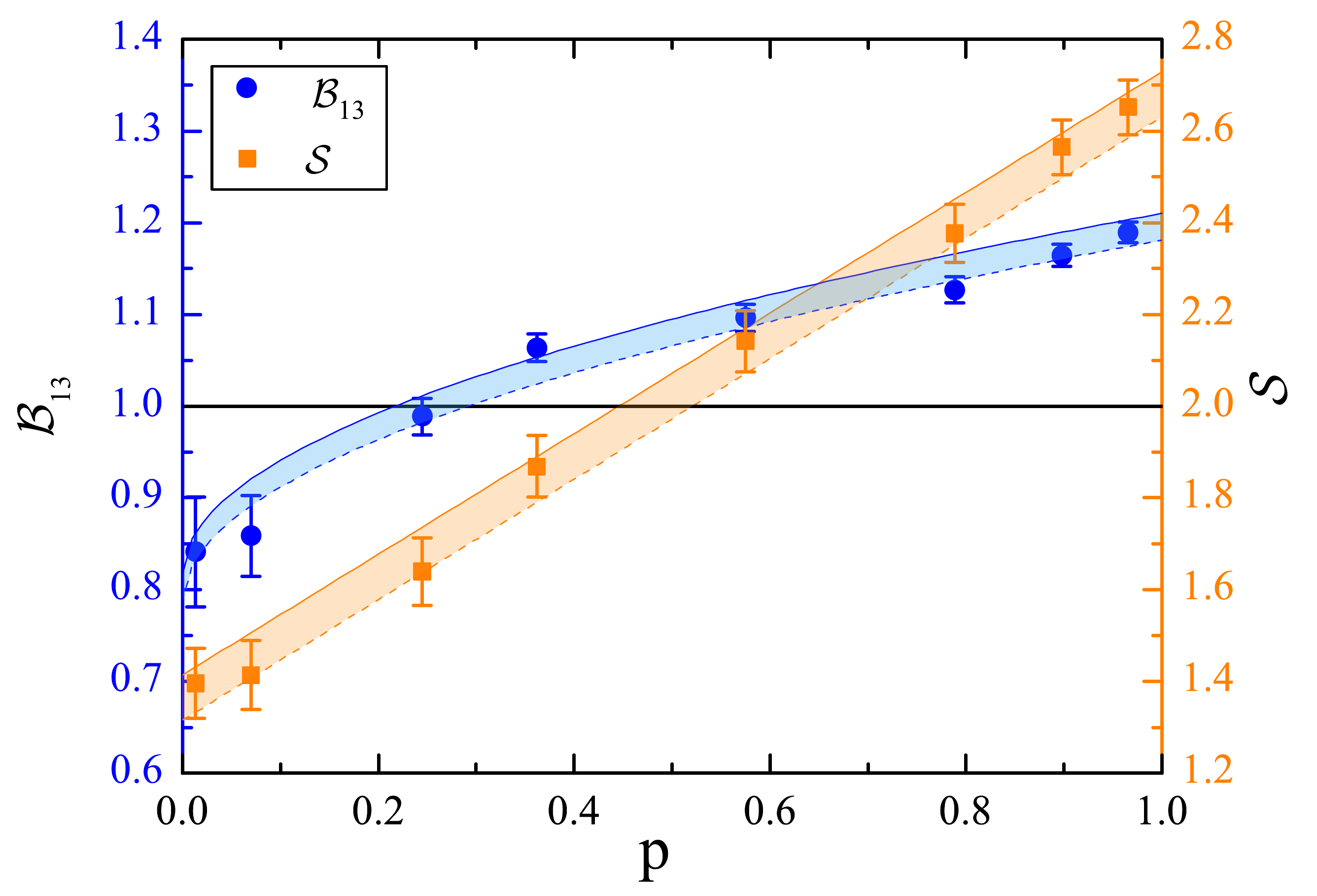}
\caption{\textbf{$\mathcal{B}_{13}$ and $S$ versus the noise parameter $p$.} The filled blue circles and filled orange squares are $\mathcal{B}_{13}$ and $S$  measured in the experiment, respectively. Each data point stands for a accumulation time of 8000~s in the experiment. The error bars  represent 1 standard deviation, assuming Poisson statistics. The blue and orange-shaded areas represent 1 standard deviation according to theoretical predictions (see Supplementary Information). }
\label{fig:ExpResult}
\end{figure}

We measure $\mathcal{B}_{13}=1.181\pm0.004$ in the experiment, which exceeds the bound ($\mathcal{B}_{13}\le1$) of bilocal hidden variable models by 45 standard deviations. We also measure a CHSH game value of $\mathcal{S} = 2.652\pm0.059$ in the Bell inequality test after entanglement swapping (see Supplementary Information), which exceeds the bound ($\mathcal{S}\le2$) of local hidden variable models by 11 standard deviations. Therefore, our experimental results reject both local and bilocal hidden variable models with high confidence. 
We study the response of both parameters to the influence of noise. As shown in Fig.\ref{fig:ExpResult}, as we increase the noise by delaying a single photon pulse with respect to the other in the BSM from 0 (corresponding to the noise parameter $p=1$) to a significant level ($p< \frac{1}{2}$), both the values of $\mathcal{B}_{13}$ and $\mathcal{S}$ decrease. We notice that $\mathcal{B}_{13}$ remains above 1 even at a significant noise level, where $\mathcal{S}<2$. The experimental results are consistent with theoretical results (shaded areas), and confirm the theoretical prediction that the rejection of bilocal hidden variable models is more noise tolerant (see Supplementary Information for details).

We highlight several important achievements in this experiment. By bringing the laser from spontaneous emission to stimulated emission periodically to output laser pulses with randomized phases for SPDC process in each source, we close the independent source loophole. We use the same mechanism to generate random bits for measurement setting choices. 
By separating the random bit generation space-like from the creation of entanglement in the sources, we satisfy the requirement of measurement independence in the experiment~\cite{Giustina2015LoopholeFreeBT,Shalm2015LoopholeFreeBT}. We also close the locality loophole. The detection loophole may be closed in the future with improved single photon detection efficiency. The next step may be to explore quantum networks with more advanced topological structures~\cite{Tavakoli2014StarNet,Chaves2016BellNetwork,Andreoli2017Nlocal,Gisin2017Tri} and for novel applications such as device-independent quantum information processing~\cite{Lee2018DIInfProc}.

\vspace{30pt}

\appendix

\centerline{\textbf{\large{Supplementary Information}}}

\section{Polarization entangled photon-pair sources}

In our experiment, we generate polarization entangled photon-pairs in the Bell state $\ket{\Phi^+}=\frac{1}{\sqrt{2}}(\ket{H}\ket{H}+\ket{V}\ket{V})$ via Type-0 spontaneous parametric down-conversion in periodically poled MgO doped Lithium Niobate (PPMgLN), where $\ket{H}$ and $\ket{V}$ represent horizontal and vertical polarization states, respectively. The DFB laser emits 2-ns laser pulse (central wavelength 1558~nm) at a repetition rate of 250 MHz. A 40-GHz intensity modulator (IM) is used to carve the 2-ns laser pulses into 90 ps laser pulses. Both the DFB laser and the IM are driven by a pules pattern generator (PPG). The laser pulses are amplified by an Erbium doped fibre amplifier (EDFA) and fed into a PPMgLN crystal for second harmonic generation (SHG). The SHG pulses are coupled into a 780~nm single-mode fibre. The residual pump laser pulses are highly attenuated and then output to the free space through a fibre coupler. After filtered by a 855-nm bandpass filter, the 779-nm laser pulses are used to pump a 2.5-cm-long PPMgLN crystal placed inside a polarization Sagnac loop. The focal length of the off-axis parabolic mirrors (OPM) is 101.6~mm and the beam diameter at the beam waist is 108~$\mu\mathrm{m}$. Photon-pairs are generated via the Type-0 SPDC in the PPMgLN crystal. Two dichroic mirrors (DM) are used to separate the photon-pairs from the pump pulses. After passing through a silicon plate, the photons are collected into the fibre. The photon-pairs with wavelength at 1560~nm and 1556~nm are selected using a set of dense wavelength-division multiplexing (DWDM) filters. To reduce the frequency correlation of photon-pairs, fibre Bragg gratings (FBG) with 3.3~GHz bandwidth are used to make the coherence time of photons longer than the duration of pump pulses.  Following Ref~[\cite{Ryan2010OptPDC}], by setting the beam waists to be 54~$\mu$m and 55~$\mu$m at the center of the crystal for the pump (779 nm) and daughter photons (1556 nm or 1560 nm), respectively, we achieve an overall detection efficiency for single photon from the creation to detection to be 12\%. Note that this includes the loss due to the use of narrow bandpass filters. 

\begin{figure}[!htbp]
\centering
\includegraphics[width=\textwidth]{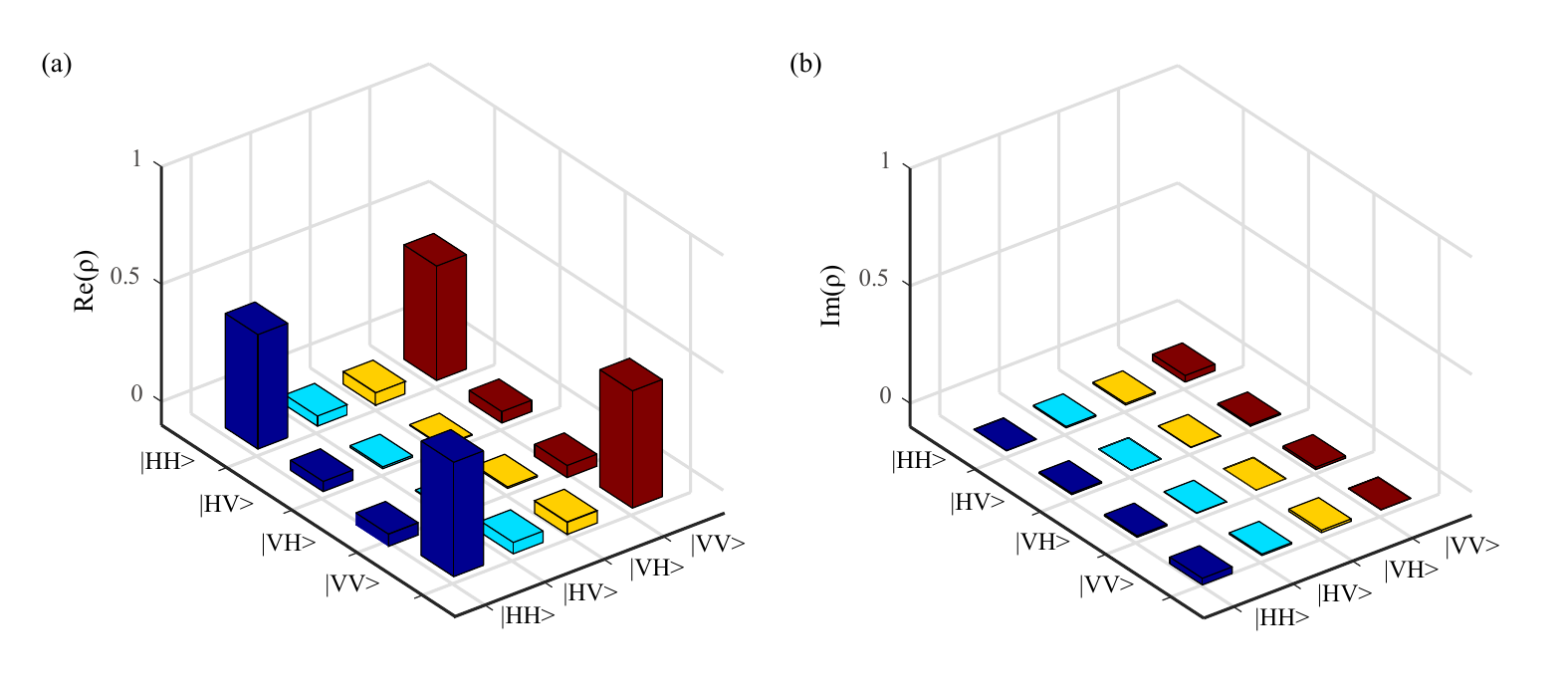}
\caption{\textbf{The result of quantum state tomography measurement.} \textbf{a} real part and \textbf{b} imaginary part of the reconstructed density matrix. }
\label{fig:dm}
\end{figure}

To characterize the generated entangled state, we perform a quantum state tomography measurement on the two-photon state\cite{James2001Tmomography}. The reconstructed density matrix is shown in Fig. \ref{fig:dm}. The fidelity ($F=\braket{\Phi^+|\rho|\Phi^+}$) of the entangled state is $0.9853\pm0.0009$, where the uncertainty is obtained using a Monte Carlo routine assuming Poissonian statistics.

Considering the excess noise, we assume that the produced state is a Werner state in the form of  $\rho_w=V\ket{\Phi^+}\bra{\Phi^+}+\frac{1-V}{4}\mathbf{I}$, where $V$ is the visibility of the quantum state and $\mathbf{I}$ is the identity matrix. The visibility is related to the fidelity as $V=(4F-1)/3\approx 0.9804$. In our experiment, we estimate the visibility with  $\frac{C_{max}-C_{min}}{C_{max}+C_{min}}$ , where $C_{max}$ denotes the maximum two-photons coincidence counts and the $C_{min}$ the minimum coincidence counts. We optimize the visibility if the visibility drops below 97\% in the experiment.

\section{Verification of random measurement basis choice}
\subsection{Quantum random number generation}
The quantum random number generators (QRNGs) used in our experiment are based on the quantum phase noise of a single-mode laser near the lasing threshold\cite{Qi2010QRN}. In our experiment, the quantum phase noise of the laser is measured with an active stabilized polarization insensitive Michelson interferometer \cite{Nie2015QRNG}. The QRNG outputs random bits at a repetition rate of 250~MHz.

 \begin{figure}
\centering
\includegraphics[width=\textwidth]{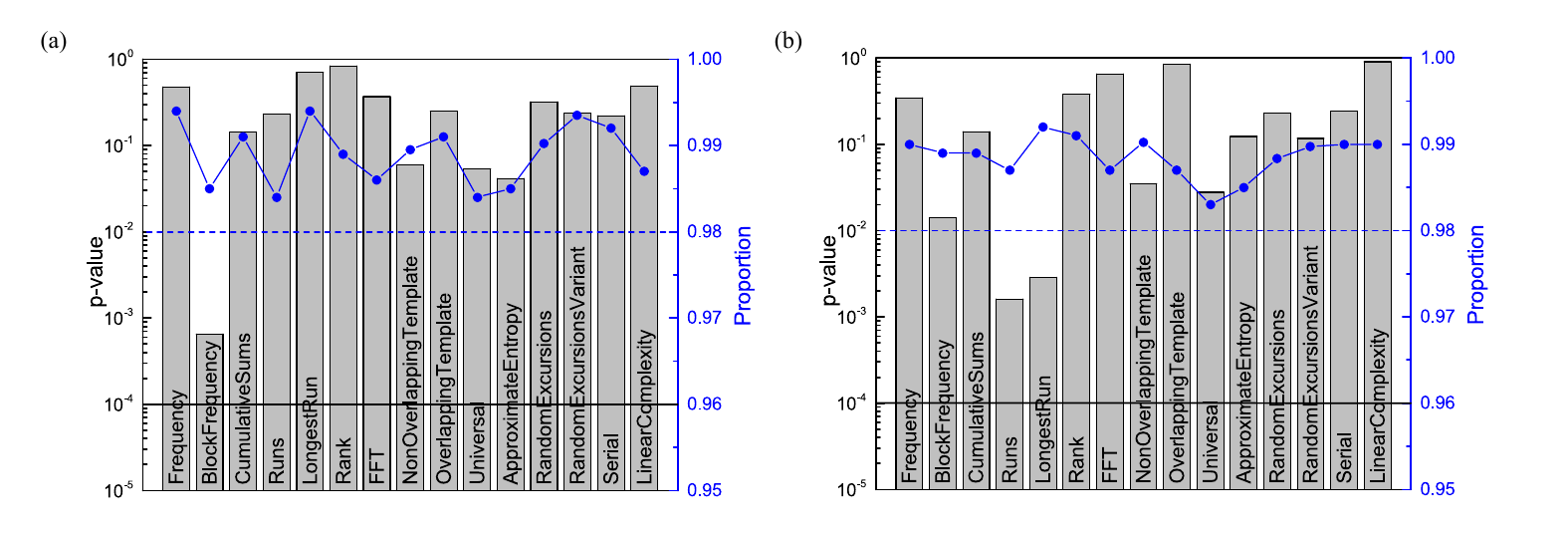}
\caption{The NIST test results of 1-Gb random bits from the QRNG of Alice (\textbf{a}) and Charlie (\textbf{b}). Both the uniformity of the p-values assessment (bars) and results for the proportion test (dots) are depicted in the figures. All the measurement results are above the thresholds represented by the solid horizontal line and dashed horizontal line for the p-value test and proportion test, respectively.}
\label{fig:qrn}
\end{figure}

A 1550-nm laser diode (LD) is driven by a constant current which is slightly above its threshold, and a thermo electric cooler (TEC) is used to stabilize its temperature. The emitted photons enter an unbalanced interferometer which converts the phase information into the intensity. The output of the interferometer is detected by a 10-GHz InGaAs photon detector (PD). The PD signals is converted to 8 bits per sample by an analog-to-digital converter (ADC) at a repetition rate of 1GHz. We choose one of the 8 bits and feed it into a field-programmable gate array (FPGA). We use the XOR algorithm which consumes 4 adjacent bits to generate 1 random bit for the measurement setting choice at a repetition rate of 250~MHz. As shown in Fig.~\ref{fig:qrn}, the generated random bits pass the NIST statistical test suite.

\subsection{Fidelity of random measurement basis choice}

In the bilocality test, one of the two measurement basis $\hat{M}_0=(\sqrt{2}\sigma_z+\sigma_x)/\sqrt{3}$ and $\hat{M}_1=(\sqrt{2}\sigma_z-\sigma_x)/\sqrt{3}$ is selected according to the received random number ``0"  or ``1", respectively. We feed photons in the eigenstate of, e.g., $\hat{M}_0$, to the measurement setup and record the detection results of the two SNSPDs if the measurement base is switched to $\hat{M}_0$. Ideally, all photons should go to the assigned part. In practice, because of device imperfection, some photons may go to the other port. We define the fidelity of random basis choice as 
\begin{equation}
F_m=\frac{C_r}{C_r+C_w},
\end{equation}
where $C_r$ represents the photons recorded by the correct SNSPD and $C_w$ the photons recorded by the wrong SNSPD, respectively. We do the same for $\hat{M}_1$. In our experiment, the fidelity measured in the two cases are  $(98.49\pm0.4)\%$ and $(99.56\pm0.03)\%$, respectively. The average fidelity is $(99.02\pm0.20)\%$.

\section{Space-time analysis of the experiment}

\begin{figure}[h]
\centering
\includegraphics[width=0.618\textwidth]{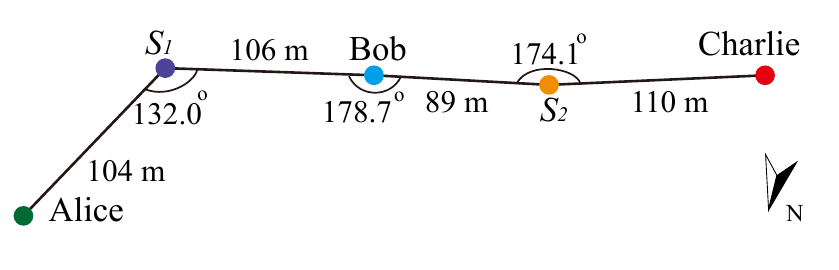}
\caption{\textbf{The relative positions of the 2 sources and 3 measurement nodes with the beeline distance.}}
\label{fig:loc}
\end{figure}

\begin {table}
\centering
\caption{Details of  the length of fibre between adjacent nodes.}
\vspace{20pt}
\begin{tabular}{|c|c|c|c|c|}

\hline
Link&Alice-$S_1$&$S_1$-Bob&Bob-$S_2$&$S_2$-Charlie\\
\hline
Length (m)&110.98&124.52&108.13&124.55\\

\hline
\end{tabular}
\label{tab:loc}
\end{table}
Strict locality constraints should be met in our experiment in order to close the locality loophole and independent sources loophole. The relative positions of the 2 sources and 3 measurement nodes in our experiment are shown in Fig.~\ref{fig:loc}. The locality constraints set four space-like separation conditions in each experimental trial.

(1) Space-like separation between the two state emission events in sources $S_{1}$ and $S_{2}$ requires $L_{12}/c >\mathrm{max}(t_{12}+\tau_1, t_{12}+\tau_2)$, where $L_{12}$ is the beeline distance between $S_{1}$ and $S_{2}$, $t_{12}$ is the relative delay between the earliest time of the state emission (the creation of pump laser pulses in the two sources), which are also taken as the earliest time to create the local hidden variables,  $\tau_1$ ($\tau_2$) is the time elapse starting from the earliest time of generating a pump laser pulse to the latest time of loading the photon to the optical fibre for entanglement distribution.
 
(2) Space-like separation between the event of generating quantum random numbers for Alice (QRNGA) and state emission events in sources $S_{1}$ and $S_{2}$, respectively requires $L_{A1}/c > (t_{Aq1}+\tau_{Aq})$ and $L_{A2}/c > (t_{Aq2}+\tau_{Aq})$, where $L_{A1}$ ($L_{A2}$) is the beeline distance between Alice and $S_{1}$ ($S_{2}$), $t_{Aq1}$ ($t_{Aq2}$) is the relative delay between the earliest time of quantum random number generation and the earliest time of state emission in sources, and $\tau_{Aq}$ is the time elapse for QRNGA to output a random bit. Similarly, space-like separation between the quantum random number generation event for Charlie (QRNGC) and state emission events in sources $S_{1}$ and $S_{2}$ requires $L_{C1}/c > (t_{Cq1}+\tau_{Cq})$ and $L_{C2}/c > (t_{Cq2}+\tau_{Cq})$, where $L_{C1}$ ($L_{C2}$) is the beeline distance between Charlie and $S_{1}$ ($S_{2}$), $t_{Cq1}$ ($t_{Cq2}$) is the relative delay between the earliest time for quantum random number generation and the earliest time of state emission in sources, and $\tau_{Cq}$ is the time elapse for QRNGC to output a random bit.

(3) Space-like separation between QRNGA and QRNGC requires $L_{AC}/c > \mathrm{max}(t_{AqCq}+\tau_{Aq}, t_{AqCq}+\tau_{Cq})$, where $L_{AC}$ is the distance between node Alice and node Charlie, and $t_{AqCq}$ is the relative delay between the earliest time of QRNGA and that of QRNGC.

(4) Space-like separation between QRNGA  and the measurement events by Bob and Charlie requires $L_{AB}/c>t_{AqB}+\tau_{B}$ and $L_{AC}/c>t_{AqC}+\tau_{C}$, where $L_{AB}$ and $L_{AC}$ denote the beeline distance between the nodes, $t_{AqB}$  and $t_{AqC}$ denote the relative delay between the earliest time of QRNGA and measurement events of Bob and Charlie, respectively, and $\tau_B$ and $\tau_C$ denote the time elapse for the measurement events, which is the interval between the time when a photon enters the loop interferometer for polarization measurement and the time when the SNSPD outputs a signal. Similarly, space-like separation between QRNGC and the measurement events by Bob and Alice requires
$L_{CB}/c>t_{CqB}+\tau_{B}$ and $L_{CA}/c>t_{CqA}+\tau_{A}$, where $L_{CB}$ and $L_{CA}$ denote the beeline distance between the nodes, $t_{CqB}$ and $t_{CqA}$ denote the relative delay between the earliest time of QRNGC and measurement events of Bob and Alice, respectively, and $\tau_B$ and $\tau_A$ denote the time elapse for the measurement events, which is the interval between the time when a photon enters the loop interferometer for polarization measurement and the time when the SNSPD outputs a signal.

The four sets of space-like separation criteria are analysed using the parameters aforementioned. We measure the relevant beeline distance, fibre length between the nodes, and the relevant elapsed time with the beeline distances and their relative angles shown in Fig.~\ref{fig:loc} and the fibre length between adjacent nodes shown in Table.~\ref{tab:loc}. The angles in Fig.~\ref{fig:loc}  are calculated using the beeline distances. In Table.~\ref{tab:spacetime}, the difference between the LHS and RHS of the inequalities are given showing that the locality and independent sources  loopholes are closed in our experiment. 

\begin{table}
\centering
\caption{Details of the space-like separation condition. The subscripts, ``1" and ``2" represent $S_1$ and $S_2$;
A, B, and C represent nodes A, B, and C; Aq and Cq represent the quantum random number in nodes A and C, respectively.}
\label{tab:spacetime}
\vspace{15pt}
\scalebox{0.9}[0.9]{
\begin{tabular}{|c|c|c|c|c|c|} 
\hline
Space-like separation conditions&\multicolumn{2}{|c|}{Beeline distance (m)}&\multicolumn{2}{|c|}{Time (ns)}&Difference (ns)\\  
\hline
\multirow{3}{*}{$L_{12}/c>max(t_{12}+\tau_{1}, t_{12}+\tau_{2})$}&\multirow{3}{*}{$L_{12}$}&\multirow{3}{*}{195}&$ t_{12}$&74.3&\multirow{3}{*}{415.50}\\
\cline{4-5}
\multirow{3}*{}&\multirow{3}*{}&\multirow{3}*{}&$\tau_{1}$&160.2&\multirow{3}*{}\\
\cline{4-5}
\multirow{3}*{}&\multirow{3}*{}&\multirow{3}*{}&$\tau_{2}$&154.4&\multirow{3}*{}\\
\hline
\multirow{2}{*}{$L_{A1}/c> t_{Aq1}+\tau_{Aq}$}&\multirow{2}{*}{$L_{A1}$}&\multirow{2}{*}{104}&$ t_{Aq1}$&266.7&\multirow{2}{*}{44.47}\\
\cline{4-5}
\multirow{2}*{$L_{A2}/c> t_{Aq2}+\tau_{Aq}$}&\multirow{2}*{$L_{A2}$}&\multirow{2}*{277}&$ t_{Aq2}$&192.4&\multirow{2}*{695.43}\\
\cline{4-5}
\multirow{2}*{}&\multirow{2}*{}&\multirow{2}*{}&$\tau_{Aq}$&35.5&\multirow{2}*{}\\
\hline

\multirow{2}{*}{$L_{C1}/c> t_{Cq1}+\tau_{Cq}$}&\multirow{2}{*}{$L_{C1}$}&\multirow{2}{*}{305.5}&$ t_{Cq1}$&360.0&\multirow{2}{*}{622.83}\\
\cline{4-5}
\multirow{2}*{$L_{C2}/c>t_{Cq2}+\tau_{Cq}$}&\multirow{2}*{$L_{C2}$}&\multirow{2}*{110}&$t_{Cq2}$&286.7&\multirow{2}*{44.47}\\
\cline{4-5}
\multirow{2}*{}&\multirow{2}*{}&\multirow{2}*{}&$\tau_{Cq}$&35.5&\multirow{2}*{}\\
\hline
\multirow{3}{*}{$L_{AC}/c>max( t_{AqCq}+\tau_{Aq}, t_{AqCq}+\tau_{Cq})$}&\multirow{3}{*}{$L_{AC}$}&\multirow{3}{*}{384.2}&$ t_{AqCq}$&94.3&\multirow{3}{*}{1150.87}\\
\cline{4-5}
\multirow{3}*{}&\multirow{3}*{}&\multirow{3}*{}&$\tau_{Aq}$&35.5&\multirow{3}*{}\\
\cline{4-5}
\multirow{3}*{}&\multirow{3}*{}&\multirow{3}*{}&$\tau_{Cq}$&35.5&\multirow{3}*{}\\
\hline
\multirow{2}{*}{$L_{AB}/c> t_{AqB}+\tau_{B}$}&\multirow{2}{*}{$L_{AB}$}&\multirow{2}{*}{191.8}&$ t_{AqB}$&496.0&\multirow{2}{*}{88.23}\\
\cline{4-5}
\multirow{2}*{$L_{CB}/c> t_{CqB}+\tau_{B}$}&\multirow{2}{*}{$L_{CB}$}&\multirow{2}*{199}&$ t_{CqB}$&401.7&\multirow{2}*{206.53}\\
\cline{4-5}
\multirow{2}*{}&\multirow{2}*{}&\multirow{2}{*}{}&$\tau_{B}$&55.1&\multirow{2}*{}\\
\hline
\multirow{2}{*}{$L_{AC}/c> t_{AqC}+\tau_{C}$}&\multirow{2}{*}{$L_{AC}$}&\multirow{2}{*}{384.2}&$ t_{AqC}$&576.2&\multirow{2}{*}{637.67}\\
\cline{4-5}
\multirow{2}*{$L_{AC}/c> t_{CqA}+\tau_{A}$}&\multirow{2}{*}{}&\multirow{2}*{}&$\tau_C$&66.8&\multirow{2}*{879.77}\\
\cline{4-5}
\multirow{2}*{}&\multirow{2}*{}&\multirow{2}{*}{}&$t_{CqA}$&335.6&\multirow{2}*{}\\
\cline{4-5}
\multirow{2}*{}&\multirow{2}*{}&\multirow{2}{*}{}&$\tau_{A}$&65.3&\multirow{2}*{}\\
\hline

\end{tabular}
}
\end{table}

\section{Synchronization and calibration} 
\begin{figure}[h]
\centering
\includegraphics[width=1\textwidth]{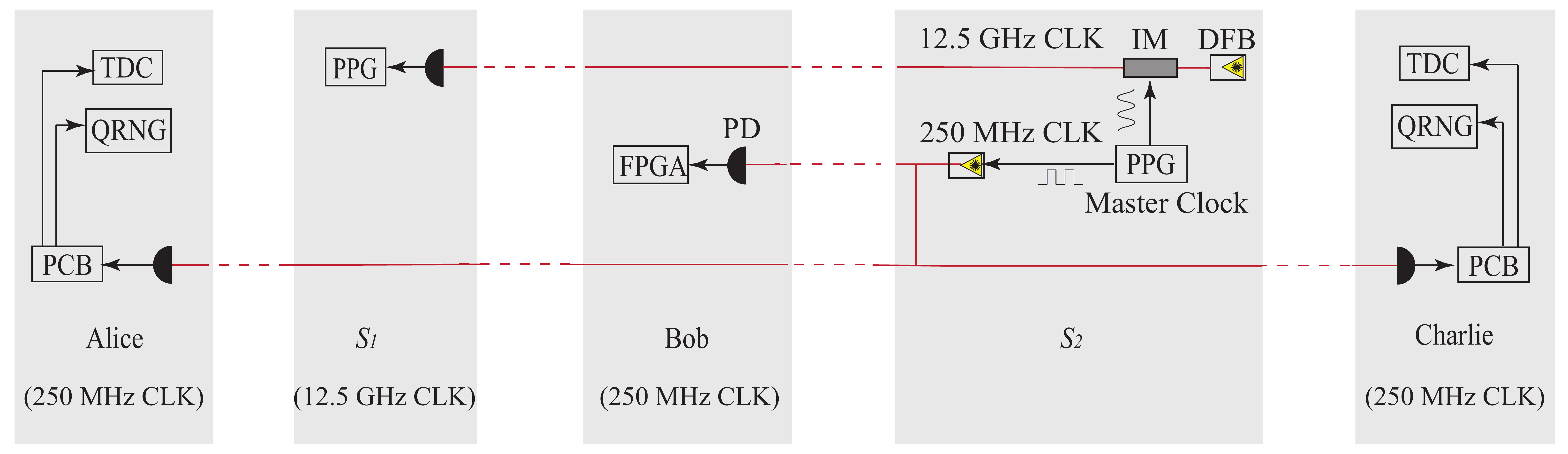}
\caption{Schematics for the experimental synchronization. The red dashed lines and black lines represent the fibre channels and coaxial cables, respectively.}
\label{fig:syn}
\end{figure}

The scheme for the synchronization of our experimental system is shown in Fig.~\ref{fig:syn}. The pulse pattern generator (PPG) in source $S_2$ is the master clock, which generates a 12.5-GHz sinusoidal signal to synchronize the two quantum sources and a 250-MHz square wave signal to synchronize the operation and measurement at the three measurement nodes.

The 12.5-GHz sinusoidal signal is used to drive an intensity modulator (IM) so that the 1550~nm CW laser emitted by DFB laser diode is carved into 12.5-GHz laser pulses. This laser pulse is sent to $S_1$ through optical fibre and detected by a 10-GHz PD. The relative low bandwidth of the PD results in a low detection signal amplitude. The detection signal is amplified with a 40-GHz microwave amplifier and then used as the external clock signal for the PPG in $S_1$.

To measure the time jitter between the pump laser pulses of the two sources, we send both of them to Bob and analyse the relative delay between them using a sampling oscilloscope. The root mean square (RMS) value of the time jitter between the two pump lasers is about 4~ps, which is much smaller than the 133-ps coherent time of the single photons.

The 250-MHz square wave signal is used to trigger an electrically driven laser diode. The generated laser pulses are sent to the three measurement nodes where they are detected by 10-GHz PDs. In Alice and Charlie's nodes, the detection signals of the PDs are fed to home-made printed circuit boards (PCBs) to generate trigger signals for the QRNGs and the time-to-digital converters (TDCs). In Bob's node, the detection signals are used to trigger the FPGA. The RMS time jitter between the single photons and the measurement/operation clock in the three nodes is about 100~ps.

\section{Bell state measurement and Hong-Ou-Mandel interference Entanglement swapping }
\begin{figure}[h]
\centering
\includegraphics[width=0.6\textwidth]{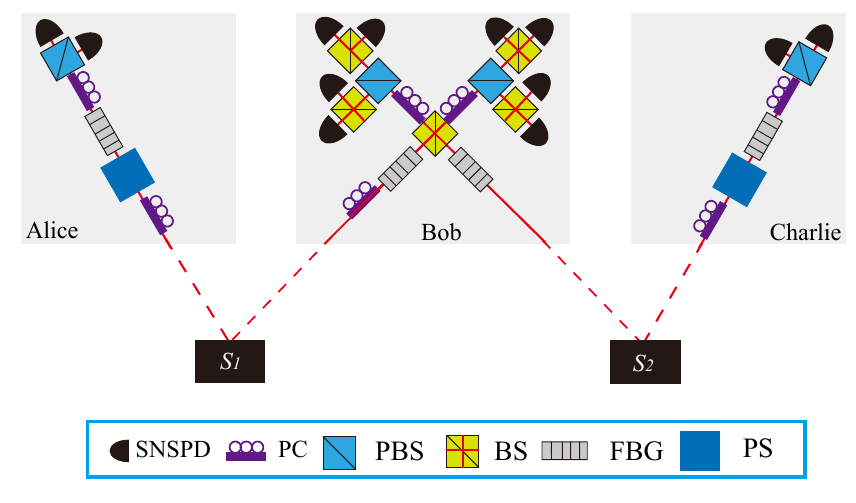}
\caption{\textbf{Schematics of experimental setup for HOM interference and entanglement swapping.}}
\label{fig:hom1}
\end{figure}

We use the experimental setup as shown in Fig.~\ref{fig:hom1} to realize entanglement swapping. Each quantum source generates entangled photon-pairs in quantum state $\ket{\Phi^\pm}=\frac{1}{\sqrt{2}}(\ket{H}\ket{H}\pm\ket{V}\ket{V})$.  The product state of the entangled photon-pairs from the two sources is given by
\begin{equation}
\begin{split}
\ket{\Phi}_{ABB^\prime C}=&\frac{1}{2}(\ket{\Phi^+}_{AC}\ket{\Phi^+}_{BB^\prime}+\ket{\Phi^-}_{AC}\ket{\Phi^-}_{BB^\prime}+\\&\ket{\Psi^+}_{AC}\ket{\Psi^+}_{BB^\prime}+\ket{\Psi^-}_{AC}\ket{\Psi^-}_{BB^\prime}),
\end{split}
\end{equation}
where $A$ ($C$) denotes the photon held by Alice (Charlie) and $B$ ($B{^\prime}$) denotes the photon held by Bob. Bob randomly projects the received photons ($B$ and $B{^\prime}$) onto one of the four Bell states with an equal probability ($25\%$) in the Bell state measurement (BSM). Correspondingly, the photons $A$ and $C$ are projected onto the same Bell state.

As shown in Fig.~\ref{fig:hom1}, we realize the BSM using the beam splitters (BSs) and polarizing beam splitters (PBSs) at Bob's node. The single photons from the two sources interference on the first BS. The two photons in state $\ket{\Psi^-}$ exit from different ports of the BS. If the two photons exit the same port of the BS, the two photons in state $\ket{\Psi^+}$ exit from different ports of the PBS, and the photons in state $\{\ket{\Phi^-}\mathrm{or}\ket{\Phi^+}\}$ exit from the same port and the photon number information can be resolved with a BS with 50\% of success. The photons are detected by eight SNSPDs and the detection results are processed by the FPGA.

\begin{figure}[h]
\centering
\includegraphics[width=0.6\textwidth]{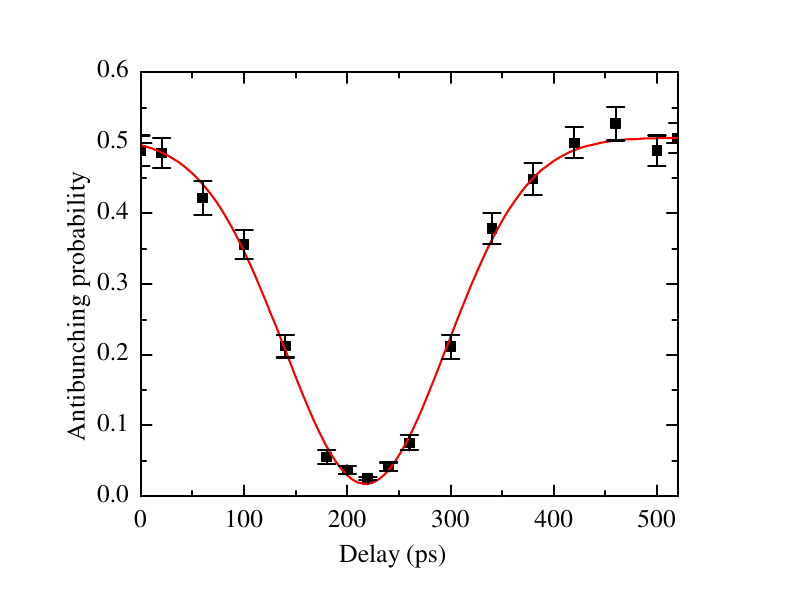}
\caption{\textbf{Experiment result of HOM interference.} Each data point is accumulated for more than 600~s. The visibility of the fitted curve is $(96.5\pm1.6)\%$.}
\label{fig:hom2}
\end{figure}

Shown in Fig.~\ref{fig:hom2} is the Hong-Ou-Mandel measure by Bob with photons from the two separate sources. By suppressing the distinguishability of photons in spectral, spatial, temporal, and polarization modes, we obtain a fitted visibility of $(96.5\pm1.6)\%$.

The imperfect of visibility is mainly due to the multi-photon pair events. For an average photon pair number per pulse $\mu\approx0.012$ in our experiment, the visibility is limited by an upper bound~\cite{Fulconis2007QIntPhonCryFib}
\begin{equation}
\mathcal V_{HOM,Max} \approx {\frac{1+8\mu}{1+12\mu}}\approx95.8\%.
\end{equation}
To reduce the contributions from the multi-photon-pair events, we discard the events when more than two SNSPDs click simultaneously at Bob's node. The upper bound is improved to
\begin{equation}
\mathcal V_{HOM,Max} \approx {\frac{1+4\mu}{1+6\mu}}\approx97.8\%.
\end{equation}

\section{Experimental violation of CHSH inequality and bilocal inequalities}
In our experiment, we examine the bilocal inequality of $\mathcal{B}_{13}\leq1$. Bob's outputs are \textbf{b}=00, 01, and \{10 or 11\} corresponding to the state projections of $\ket{\Psi^+}$ and $\ket{\Psi^-}$, and the group of $\{\ket{\Phi^+} \mathrm{or} \ket {\Phi^-}\}$ in BSM. With the measurement setting choice of $\hat{A}_0=\hat{C}_0=(\sqrt{2}\sigma_z+\sigma_x)/\sqrt{3}$ and $\hat{A}_1=\hat{C}_1=(\sqrt{2}\sigma_z-\sigma_x)/\sqrt{3}$ for Alice and Charlie, respectively, we obtain $\mathcal{B}_{13}$=$1.181\pm0.004$ in our experiment.  The the probability distribution, $P_{13}(a,b,c|x,z)$, are listed in Table.\ref{tab:P}. 
\begin {table}

\caption{Measured probabilities $P_{13}(a,b,c|x,z)$in our test of $\mathcal{B}_{13}$}
\vspace{15pt}
\renewcommand{\arraystretch}{0.7}
\begin{minipage}[t]{0.50\textwidth}

\scalebox{0.9}[0.9]{
\begin{tabular}{ccccc|c}

$x$&$z$&$\mathbf{b}$&$a$&$c$&$P_{13}(a,\mathbf{b},c|x,z)$\\
\hline
0&0&00&0&0&0.00333$\pm$0.00045\\
0&0&00&0&1&0.11977$\pm$0.00222\\
0&0&00&1&0&0.12510$\pm$0.00273\\
0&0&00&1&1&0.00435$\pm$0.00044\\
0&0&01&0&0&0.04124$\pm$0.00156\\
0&0&01&0&1&0.08950$\pm$0.00192\\
0&0&01&1&0&0.08780$\pm$0.00232\\
0&0&01&1&1&0.03819$\pm$0.00131\\
0&0&{10 or 11}&0&0&0.16530$\pm$0.00391\\
0&0&{10 or 11}&0&1&0.03899$\pm$0.00175\\
0&0&{10 or 11}&1&0&0.04941$\pm$0.00245\\
0&0&{10 or 11}&1&1&0.23702$\pm$0.00395\\
\end{tabular}
}


$\left\{
\begin{aligned}
<A_0B^0C_0>_{P_{13}}=-0.64897\pm0.00812&\\
<A_0B^1C_0>_{P_{13}}=-0.13932\pm0.00510\\
 \end{aligned}
\right.$

\end{minipage}
\hfil
\begin{minipage}[t]{0.40\textwidth}
\scalebox{0.9}[0.9]{
\begin{tabular}{ccccc|c}

$x$&$z$&$\mathbf{b}$&$a$&$c$&$P_{13}(a,\mathbf{b},c|x,z)$\\
\hline
0&1&00&0&0&0.03955$\pm$0.00154\\
0&1&00&0&1&0.09288$\pm$0.00198\\
0&1&00&1&0&0.08018$\pm$0.00225\\
0&1&00&1&1&0.03953$\pm$0.00137\\
0&1&01&0&0&0.00509$\pm$0.00056\\
0&1&01&0&1&0.12502$\pm$0.00230\\
0&1&01&1&0&0.11935$\pm$0.00270\\
0&1&01&1&1&0.00544$\pm$0.00051\\
0&1&{10 or 11}&0&0&0.16795$\pm$0.00398\\
0&1&{10 or 11}&0&1&0.04131$\pm$0.00183\\
0&1&{10 or 11}&1&0&0.04611$\pm$0.00240\\
0&1&{10 or 11}&1&1&0.23073$\pm$0.00399\\
\end{tabular}
}
$\left\{
\begin{aligned}
<A_0B^0C_1>_{P_{13}}=-0.64369\pm0.00820&\\
<A_0B^1C_1>_{P_{13}}=+0.14071\pm0.00514\\
 \end{aligned}
\right.$
\end{minipage}

\begin{minipage}[t]{0.50\textwidth}
\scalebox{0.9}[0.9]{
\begin{tabular}{ccccc|c}

$x$&$z$&$\mathbf{b}$&$a$&$c$&$P_{13}(a,\mathbf{b},c|x,z)$\\
\hline
1&0&00&0&0&0.04083$\pm$0.00156\\
1&0&00&0&1&0.09202$\pm$0.00196\\
1&0&00&1&0&0.08223$\pm$0.00226\\
1&0&00&1&1&0.00333$\pm$0.00045\\
1&0&01&0&0&0.00494$\pm$0.00055\\
1&0&01&0&1&0.12446$\pm$0.00227\\
1&0&01&1&0&0.11786$\pm$0.00266\\
1&0&01&1&1&0.00333$\pm$0.00045\\
1&0&{10 or 11}&0&0&0.17218$\pm$0.00398\\
1&0&{10 or 11}&0&1&0.00333$\pm$0.00045\\
1&0&{10 or 11}&1&0&0.04624$\pm$0.00238\\
1&0&{10 or 11}&1&1&0.03937$\pm$0.00177\\
\end{tabular}
}
$\left\{
\begin{aligned}
<A_1B^0C_0>_{P_{13}}=-0.65453\pm0.00814&\\
<A_1B^1C_0>_{P_{13}}=+0.13554\pm0.00508\\
 \end{aligned}
\right.$
\end{minipage}
\hfil
\begin{minipage}[t]{0.40\textwidth}
\scalebox{0.9}[0.9]{
\begin{tabular}{ccccc|c}

$x$&$z$&$\mathbf{b}$&$a$&$c$&$P_{13}(a,\mathbf{b},c|x,z)$\\
\hline
1&1&00&0&0&0.00513$\pm$0.00056\\
1&1&00&0&1&0.12721$\pm$0.00231\\
1&1&00&1&0&0.12126$\pm$0.00045\\
1&1&00&1&1&0.00378$\pm$0.00042\\
1&1&01&0&0&0.04046$\pm$0.00271\\
1&1&01&0&1&0.09689$\pm$0.00202\\
1&1&01&1&0&0.08241$\pm$0.00228\\
1&1&01&1&1&0.03289$\pm$0.00123\\
1&1&{10 or 11}&0&0&0.16752$\pm$0.00397\\
1&1&{10 or 11}&0&1&0.03508$\pm$0.00168\\
1&1&{10 or 11}&1&0&0.04231$\pm$0.00230\\
1&1&{10 or 11}&1&1&0.24507$\pm$0.00403\\
\end{tabular}
}
$\left\{
\begin{aligned}
<A_1B^0C_1>_{P_{13}}=-0.68071\pm0.00815&\\
<A_1B^1C_1>_{P_{13}}=-0.13362\pm0.00514\\
 \end{aligned}
\right.$
\end{minipage}
\label{tab:P}
\end{table}

We also examine the CHSH inequality conditioned on the BSM outcome $\ket{\Psi^-}$. Alice and Charlie randomly set the measurement basis with $\hat{A}_0=\sigma_z$ or $\hat{A}_1=\sigma_x$ and $\hat{C}_0=(\sigma_z+\sigma_x)/\sqrt{2}$ or $\hat{C}_1=(\sigma_z-\sigma_x)/\sqrt{2}$. We obtain the CHSH value to be $\mathcal{S}=2.652\pm0.059$.

For a study of the bilocal inequality and CHSH inequality in the presence of noise, we adjust the delay between the two photons in the BSM~\cite{Carvacho2016BellNetwork} and introduce a noise parameter

\begin{equation}
 p=\frac{C_{D}-C(d)}{C_{D}},
\end{equation}
where $C(d)$ and $C_{D}$ are two-photon coincidence events for a delay of $d$ and a complete distinguishability, respectively. The maximum value of the $p$ equals the visibility of  HOM interference, $p_{max}=(96.5\pm1.6)\%$. 

The perfect BSM for $\mathcal{B}_{13}$ can be described by the operator
\begin{equation}
 B^y=a_1\left|\Psi ^{-}\right\rangle{\left\langle\Psi ^{-}\right|}+a_2\left|\Psi ^{+}\right\rangle{\left\langle\Psi ^{+}\right|}+a_3(\left|\Phi ^{+}\right\rangle{\left\langle\Phi ^{+}\right|}+\left|\Phi ^{-}\right\rangle{\left\langle\Phi ^{-}\right|}),
\end{equation} 
with $a_1=1-2y$, $a_2=1$, $a_3=y-1$.
 Note that, the imperfect  interference  induces a mixture of $\left|\Psi ^{+}\right\rangle$ with $\left|\Psi ^{-}\right\rangle$ or $\left|\Phi ^{+}\right\rangle$ with $\left|\Phi ^{-}\right\rangle$ in the BSM, which leads to a generalized BSM operation 
\begin{equation}
B^y=a_1F_1+a_2F_2+a_3F_3,
\end{equation}
with
\begin{equation}
\begin{aligned}
F_1(p)&={\frac{1+p}{2}}\left|\Psi ^{-}\right\rangle{\left\langle\Psi ^{-}\right|}+{\frac{1-p}{2}}\left|\Psi ^{+}\right\rangle{\left\langle\Psi ^{+}\right|},\\
F_2(p)&={\frac{1+p}{2}}\left|\Psi ^{+}\right\rangle{\left\langle\Psi ^{+}\right|}+{\frac{1-p}{2}}\left|\Psi ^{-}\right\rangle{\left\langle\Psi ^{-}\right|},\\
F_3(p)&={\left|\Phi ^{+}\right\rangle{\left\langle\Phi ^{+}\right|}+\left|\Phi ^{-}\right\rangle{\left\langle\Phi ^{-}\right|}}.\\
 \end{aligned}
\end{equation}
The correlators are defined as~\cite{Branciard2012Bilocal},
\begin{equation}
\mathcal <A_xB^yC_z>=\sum_{a,b_0b_1,c}(-1)^{a+b_y+c}P(a,b_0b_1,c|x,z).
\end{equation}
According to the quantum mechanics, the measurement outcome probabilities $P(a,b_0b_1,c|x,z)$ can be written as 
\begin{equation}
P(a,b_0b_1,c|x,z) =Tr[(P_a^{x}\otimes P_{b_0b_1}\otimes P_c^{z})\cdot (\rho _{AB}\otimes \rho _{BC})].
\end{equation}
We have,
\begin{equation}
\mathcal <A_xB^yC_z>=Tr[(A_x\otimes B\otimes C_z)\cdot (\rho _{AB}\otimes \rho _{BC})].
\end{equation}
In our swapping experiment, we consider two different kinds of noise in quantum sources,\\
(1) White noise:
\begin{equation}
\mathcal  \rho=v\left|\Phi ^{+}\right\rangle{\left\langle\Phi ^{+}\right|}+{\frac{1-v}{4}}\mathbb{I}.
\end{equation}
(2) Colour noise:
\begin{equation}
\mathcal \rho=v\left|\Phi ^{+}\right\rangle{\left\langle\Phi ^{+}\right|}+{\frac{1-v}{2}}(\left|\Phi ^{+}\right\rangle{\left\langle\Phi ^{+}\right|}+\left|\Phi ^{-}\right\rangle{\left\langle\Phi ^{-}\right|}).
\end{equation}
Then, the state produced by a source can be expressed by
\begin{equation}
\mathcal \rho_{XB}=v_X\left|\Phi^+\right\rangle{\left\langle\Phi ^{+}\right|}+(1-v_X
)[\frac{\lambda_X}{2}(\left|\Phi ^{+}\right\rangle{\left\langle\Phi^{+}\right|}+\left|\Phi ^{-}\right\rangle{\left\langle\Phi ^{-}\right|})+\frac{1-\lambda_X}{4}\mathbb{I}],
\label{eq:rhoxb}
\end{equation}
where, $X=A$ or $B$, $v_X$ denotes the total noise, and $\lambda_X $ denotes the fraction of coloured noise.

We briefly introduce the derivation of $\mathcal{B}_{13}$ and $\mathcal{S}$ below. 

(1) For $B_{13}$, we have 
\begin{equation}
\begin{aligned}
\mathcal <A_xB^{0}C_z>_{P_{13}}&=\sum_{a,c}(-1)^{a+c}[P_{13}(a,00,c|x,z)+\\ &P_{13}(a,01,c|x,z)-P_{13}(a,{10\ \mathrm{or}\  01},c|x,z)],\\
\mathcal <A_xB^{1}C_z>_{P_{13},b^{0}=0}&=\sum_{a,c}(-1)^{a+c}[P_{13}(a,00,c|x,z)-P_{13}(a,01,c|x,z)].
\end{aligned}
\end{equation}
The correlators are defined as
\begin{equation}
\begin{aligned}
\mathcal I&=\frac{1}{4}\sum_{x,z}{<A_xB^{0}C_z>_{P_{13}}},\\
\mathcal J&=\frac{1}{4}\sum_{x,z}(-1)^{x+z}{<A_xB^{1}C_z>_{P_{13},b^{0}=0}},\\
\mathcal B_{13}&=\sqrt[]{\left|I\right|}+\sqrt[]{\left|J\right|}.
\end{aligned}
\end{equation}
Using the same measurement basis in our experiment, we obtain
\begin{equation}
\mathcal B_{13}=\frac{\sqrt{\left|2[v_A(1-\lambda_A)+\lambda_A][v_C(1-\lambda_C)+\lambda_C]\right|}}{\sqrt{3}}+\frac{\sqrt{pv_Av_C}}{\sqrt{6}}
\label{eq:b13povm}
\end{equation}

(2) For $\mathcal{S}$ conditioned on the BSM outcome $\ket{\Psi^-}$,  we have
\begin{equation}
\begin{aligned}
\mathcal S=\left|<A_0C_0>+<A_0C_1>+<A_1C_0>-<A_1C_1>\right|\\
=\sqrt[]{2}(pv_Av_C+\left|v_A(-1+\lambda_A)-\lambda_A)(v_C(-1+\lambda_C)-\lambda_C)\right|),
\end{aligned}
\label{eq:CHSH}
\end{equation}

In our experiment, we have $v_A\approx v_C\approx \sqrt[]{v}\approx\sqrt[]{0.93}$, here $v$ denotes the visibility of the swapping. Assuming that $\lambda_A=\lambda_C=\lambda$, we obtain the shadow areas shown in Fig.~4 of the main text, With the upper bound given by $\lambda=1$ (solid lines), and the lower bound for $\lambda=0$ (dashed lines).





\end{document}